\documentclass{ieeeaccess}
\usepackage{cite}
\usepackage{amsmath,amssymb,amsfonts}
\usepackage{algorithmic}
\usepackage{graphicx}
\usepackage{textcomp}
\usepackage{booktabs}
\usepackage{array}
\usepackage{color}
\usepackage{amsmath}
\usepackage{amsfonts}
\usepackage{multirow}
\usepackage{nomencl}
 \usepackage{subcaption}
 
\def\BibTeX{{\rm B\kern-.05em{\sc i\kern-.025em b}\kern-.08em
    T\kern-.1667em\lower.7ex\hbox{E}\kern-.125emX}}
\begin{document}
\history{Date of publication xxxx 00, 0000, date of current version xxxx 00, 0000.}
\doi{10.1109/ACCESS.2017.DOI}

\title{Saturation Line Forecasting via a Channel and Temporal Attention-based Network}

% \author{\uppercase{JUN YANG}\authorrefmark{1},
%  \uppercase{qing li}\authorrefmark{1},  \uppercase{and yixuan sun} \authorrefmark{2}}

 \author{\uppercase{Wei Wang}\authorrefmark{1, 2, 3}, \uppercase{Linchao Li}\authorrefmark{3}, 
 \uppercase{Xuwei Wang}\authorrefmark{4}, 
\uppercase{Qing Li}\authorrefmark{1,2},
 \uppercase{Shengyao Jia}\authorrefmark{1,2},
 \uppercase{RenYuan Tong}\authorrefmark{1,2}, and \uppercase{Yang Jun}\authorrefmark{1,2}}

\address[1]{National and Local Joint Engineering research center for Disaster Monitoring Technologies and Instruments, China Jiliang University, Hangzhou 310051, China}
\address[2]{Zhejiang Provincial Key Laboratory of Intelligent Manufacturing Quality Big Data Traceability Analysis and Application, China Jiliang University, Hangzhou 310051, China}
\address[3]{Zhejiang PeckerAI Technology., Ltd, Hangzhou 310018, China}
\address[4]{College of Computer Science and Technology, Zhejiang University, Hangzhou 310027, China}

\tfootnote{This work was supported by the Zhejiang Key Research and Development Program under Grant 2018C03040 and 2021C03016.}

\markboth
{Author \headeretal: Preparation of Papers for IEEE TRANSACTIONS and JOURNALS}
{Author \headeretal: Preparation of Papers for IEEE TRANSACTIONS and JOURNALS}

\corresp{Corresponding author: Qing Li (lq13306532957@163.com)}

% \corresp{Corresponding author: c (123456789@163.com).}

\begin{abstract}
Tailings ponds are places for storing industrial waste. The saturation line is the key factor of quantifying the safety of tailings pond. Existing saturation line time-series prediction methods are mainly based on statistical models or shallow machine learning models. Although these models aim to capture the time dependence of the sequence data, the channel and temporal are even unavailable in principle. In this paper, we presents a two-stage forecasting method, which embeds the channel and temporal attention into a hybrid CNN-LSTM model to predict the saturation line. In the first stage, the discrete wavelet transform (DWT) is applied to capture the refined sequence information. In the second stage, the CNN-LSTM model is utilized to learn the basic spatial and temporal features in the time series. Furthermore, the channel and temporal attention model are embedded into CNN-LSTM model to enhance the feature extracting ability in channel and temporal dimension. Consequently, our proposed model is shown to outperform classic models on five real-world datasets.
\end{abstract}


\titlepgskip=-15pt

\maketitle

% \makenomenclature
% \renewcommand{\nomname}{NOMENCLATURE}
% \printnomenclature 
% \hfill % add blank space
% \nomenclature{LSTM}{Long-Short-Term Memory}
% \nomenclature{ARIMA}{Auto-Regressive Integrated Moving Average}
% \nomenclature{CNN}{Convolutional Neural Network}
% \nomenclature{DWT}{Discrete wavelet transform}
% \nomenclature{RF}{Random Forest}
% \nomenclature{MAE}{Mean absolute error}
% \nomenclature{RMSE}{Root-mean-square error}
% \nomenclature{R^$2$}{Coefficient of determination}
% \nomenclature{RNN}{Recurrent neural network}
% \nomenclature{NO.8}{Dataset 8 of the tailings dam}
% \nomenclature{NO.13}{Dataset 13 of the tailings dam}
% \nomenclature{NO.17}{Dataset 17 of the tailings dam}
% \nomenclature{NO.21}{Dataset 21 of the tailings dam}
% \nomenclature{NO.28}{Dataset 28 of the tailings dam}
% \nomenclature{NO.33}{Dataset 33 of the tailings dam}
% \nomenclature{SVR}{Support vector regression}
% \nomenclature{DTR}{Decision tree regression} 
% \nomenclature{RFR}{Random forest regression}   
% \nomenclature{MLP}{Multilayer perception}
% \nomenclature{GRU}{Gate recurrent unit}
% \nomenclature{i_t}{Input gate}
% \nomenclature{f_t}{Forget gate}
% \nomenclature{c_t}{Memory cell}
% \nomenclature{o_t}{Output gate}
% \nomenclature{h_t}{Output}

\section{Introduction}
\label{sec:introduction}
Tailings ponds are places to store industrial waste. The tailings pond failure is ranked top 18th in the world's risk assessment \cite{b8}. Until now, more than 100 major tailings dam accidents were reported that caused significant damage from 1960 to 2022 \cite{b1}. The common methods for evaluating the safety situation of tailings ponds rely on manual observation or measurement analysis utilizing different sensors, $e.g.,$ water level senors, displacement sensors, rainfull sensors and deformation sensors. In fact, considering the variability of topography, mine construction conditions and weather condition, the situation of the tailings dam is complicated and changeable. There are typically tailings dams located in remote mountainous areas, their structure is intricate, and the problems associated with the breakage of the dams are almost nonlinear. This makes it impossible to directly observe the tailings pond's stability.
		 
When the saturation line drops by 1 meter, the safety factor of static stability increases by 0.05 or more, making it the most important factor of tailings dam stability \cite{b11}. If the saturation line is too high, the dam stability will be reduced, and even leakage, landslides, and dam failure may occur \cite{b32,b33,b34}. Hence, tailings dams' saturation lines are termed their lifelines \cite{b13}. 

It is imperative to establish an reliable model to predict the height of saturation line so as to evaluate the security situation of tailings pond. However, the prediction research of tailings pond is rare. To alleviate this problem, our goal is to propose a model that can take full advantage of deep learning to fit complex data. In more detail, utilizing the hidden information of the historical saturation line value, the value and tendency in the future can be predicted. Based on this, we propose a channel and temporal attention-based CNN-LSTM network. In our model, convolutional layers play important roles in extracting high-dimensional structure information and pass it on to the LSTM layers for learning time-series dependence. Furthermore, the channel-wise operation in the attention model is utilized for extracting the channel structure generated by the CNN model in high dimension. The temporal information is crucial since it expresses the implicit time series dependencies in a large degree. Meanwhile, the channel-wise features are enhanced by the channel-wise operation. Since the situation of the tailings dam is complex, the data sequence of the saturation line is unstable and without obvious periodic structure. The noise data contains misleading and takes up a lot of space or memory. To overcome this drawbacks, we applied the discrete wavelet transform (DWT) to decompose the saturation line into different time-frequency sequences, and then remove the noise in the decomposed data according to the rigrsure strategy. Through decomposition and reconstruction operation, the data is refined to show the effective time-series dependence. 
In this work, taking a chinese pond as the study area, the main contributions of our study are summarized as follows:

$\bullet$ Proposing an effective channel and temporal attention-based CNN-LSTM network to predict the saturation line, which achieves satisfied performance in terms of MAPE, RMSE and R$^2$. 

$\bullet$ Comparing our proposed model with different hyperparameters and with other state-of-the-art models.

$\bullet$ Conducting the ablation studies to confirm the components of our model $i.e.,$ channel and temporal attention, DWT, LSTM.

\section{Related Work}	
Recently, researchers are devoted on tailings pond monitoring \cite{b2,b14,b26,b41,b42,b43}. The researchers are mainly focusing on the stability status by monitoring data from sensors and make early-warnings in time by mathematical modeling method, image recognition method and data analysis. Huang et al. \cite{b3} conducted a tailings pond monitoring and early-warning system based on three-dimensional GIS, the response time of the safety monitoring and early warning system is less than 5 seconds.
Li et al. \cite{b5} proposed GPS means to monitor the displacement of tailings dam online. Gao et al. \cite{b6} established remote sensing interpretation using high-resolution remote sensing images. M.Necsoiu \cite{b7} used satellite radar interferometry to monitor the tailings sedimentation. D.F.Che et al. \cite{b8} assessed the risk of tailings pond by runoff coefficient, which can simultaneously determine the safety performance of multiple tailings dams. Dong et al. \cite{b10} set up the alarm system based on the cloud platform, showing good performance in real-time monitoring. Qiu et al.\cite{b11} designed a monitoring system of saturation line based on mixed programming.

Recently, with the advantages of handling almost any non-linear and linear problems whatever low- and high-dimensions, neural network and machine learning methods have been effectively composed in real-time risk analysis and evaluation \cite{b4,b9,b12,b19,b44,b45,b46,b47}. However, the role of real-time monitoring cannot be equated with early warning and forecasting. In other words, risk prediction methods could help people perceive risk before it happens. With excellent ability to process time-series, classic prediction model such as Auto-Regressive Integrated Moving Average (ARIMA) and LSTM have been used in prediction problems \cite{b21,b28,b48,b49,b50}.
 They analyze and identify the time series information of training data and give the prediction value for a few days in advance. Nevertheless, different from LSTM, the ARIMA model only gets a high score at the condition of data with linear correlation or without obvious fluctuation. With the rapid development of deep learning, the CNN and LSTM have been the most popular networks. The CNN could filter out the noise data and extract important features, achieving good performance in images, speech, and time-series \cite{b51,b52}. While the LSTM network has the ability to find the linear or nonlinear time series information from the shallow and deep network and combine it with current memory \cite{b53}. In light of this, combining LSTM with CNN may achieve better prediction performance to a large extent.
	
\section{DATA PREPARATION}
\begin{figure}
    \centering
    \includegraphics[scale=0.3]{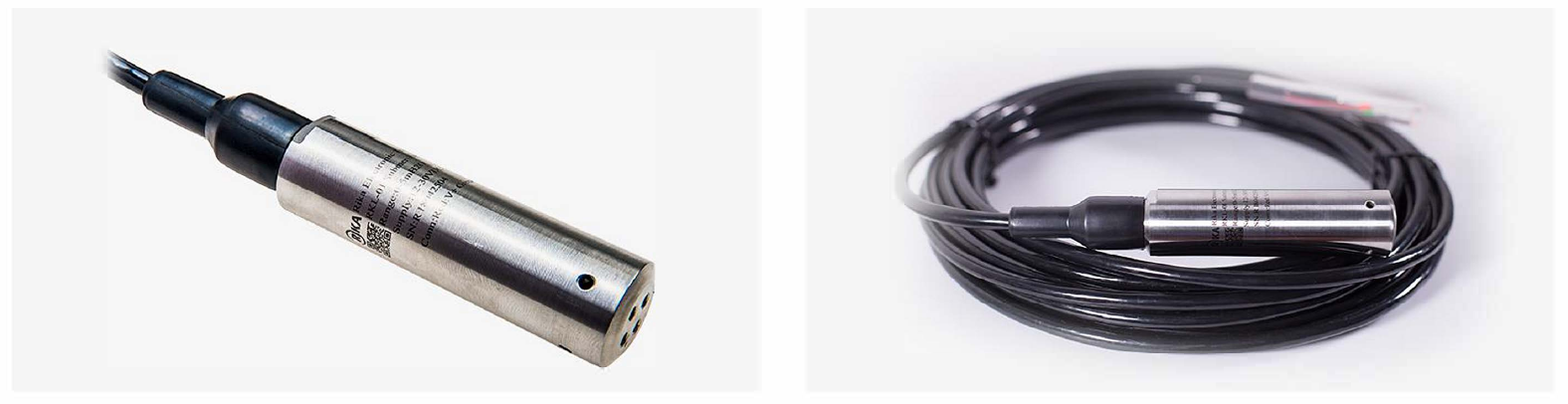}
    \caption{The water level sensors we utilized for collecting saturation line data.}
    \label{waterlevel}
\end{figure}

The study site is a copper mine tailings pond in Zhejiang Province, China. The real-world historic data for this work are collected from the water level sensors from 2020-03-18 to 2021-04-29. Specifically, we collect five datasets from different height of the tailings pond. The water level sensor is shown in Figure.~\ref{waterlevel}.

After filling the missing value with medium value and replacing the abnormal value with medium value, 9,865 data points are collected for our prediction task. The monitoring data are continuous, which maintain a wide range of original time-series information. Notably, for the 9,865 data,70\% of the dataset are set as the training sets, 10\% as the validation set. The performance of our models is confirmed on the rest 20\% dataset. We do not shuffle the data as usual in traditional deep learning studies since the time-series dependence relies on the time series order.
Table.~\ref{datapoint} indicates the describe of the monitoring data. The first three rows show the historical monitoring data, while the other columns show the statistic details.

\begin{table}
	\centering
	\caption{Describe of the datasets used for saturation line prediction.}
	\scalebox{0.95}{
	%% \tablesize{} %% You can specify the fontsize here, e.g., \tablesize{\footnotesize}.
	\begin{tabular}{ccccccccc}
	\toprule  \hline
	\textbf{}	& \textbf{NO.8}& \textbf{NO.13} &\textbf{NO.17} &\textbf{NO.21} &\textbf{NO.28} &\textbf{NO.33} \\
	\midrule
	1         &4.56 &7.78 &11.38 &10.76 &14.10 &15.01  \\
	2         &4.58 &7.79 &11.35 &10.72 &14.16 &15.07  \\
	3         &4.59 &7.81 &11.37 &10.81 &14.18 &15.27  \\
	
	\midrule
	
	mean   &4.57 &7.73 &11.32 &10.67 &14.52 &15.03 \\
	min &4.08 &6.86 &11.14 &10.08 &14.04 &14.70 \\

	25\%   &4.51 &7.68 &11.28 &11.28 &14.23 &14.96 \\
	50\%   &4.56 &7.80 &11.36 &10.76 &14.61 &15.01 \\
	75\%   &4.61 &7.84 &11.37 &10.80 &14.68 &15.04 \\
	max &4.95 &8.18 &11.40 &10.88 &15.40 &15.41  \\  \hline
	\bottomrule
	\label{datapoint}
	\end{tabular}}
\end{table}

In order to eliminate the impact of different data dimensions on the calculation, we apply \text{Z-score} normalization, the formula is as follows:
\begin{equation}
	\dot{x}=\frac{x_t-\mu_t}{\sigma_t}
\end{equation}
where $ x_t $ is the input data, $\mu_t$ and $\sigma_t$ are the averages and standard deviation of data.
		
\section{Method}
This section firstly describes the overview of our model. The DWT method is indicated in Sec.~\ref{DWT}, our proposed model is explained in Sec.~\ref{proposed}.

\begin{figure*}[tbp]
    \centering
    \includegraphics[width=1\linewidth]{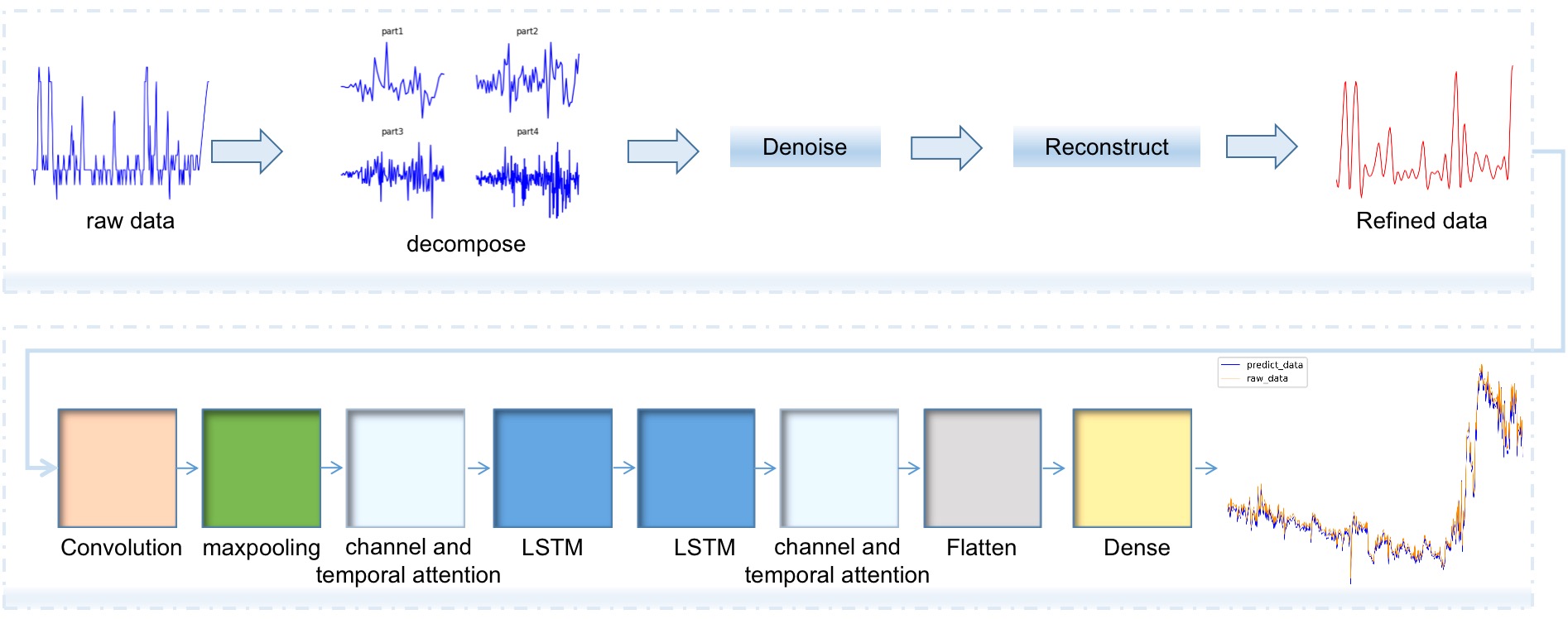}
    \caption{The Wavelet-CNN-LSTM model. The first row is the de-noise process. The de-noised data is feed to the CNN-LSTM$^2$ model (second row) for predicting. The channel and temporal attention is shown in Figure.~\ref{channel-spatial}.}
    \label{framework}
\end{figure*}

\subsection{Overview}
In this paper, we presents a two-stage forecasting method, which
embeds the channel and temporal attention into a CNN-LSTM model to predict the saturation line. In the
first stage, $i.e.$, first row in Figure.~\ref{framework}, the DWT is applied to capture the refined sequence information. In
the second stage, $i.e.$, second row in Figure.~\ref{framework}, the CNN-LSTM model is used to learn the spatial and temporal features in the refined time series. Furthermore, the channel and temporal attention model are utilized to enhance the feature extracting
ability.

\subsection{DISCRETE WAVELET TRANSFORM}
\label{DWT}
Although the window Fourier transform (short-time Fourier transform) can partially locate the time, since the window size is fixed, it is only suitable for stationary signals with small frequency fluctuations, and not suitable for non-stationary signals with large frequency fluctuations. As a signal time-frequency analysis method, the wavelet transform (WT) can automatically adjust the window size according to the frequency. What has greatly contributed the effectiveness of WT is the truth that it is an adaptive time-frequency analysis method which can perform multi-resolution analysis. As a result, wavelet transform is known as a microscope for analyzing and processing signals. In our study, we apply the discrete wavelet transform (DWT) \cite{b26} to decompose the collected saturation line data of tailings pond in to 4 frequency sequences. After removing the noise in the decomposed data, the wavelets are reconstructed to obtain new integrated data for further multi-resolution study. 

The WT refers to the displacement of a certain basic wavelet function by $\omega$ units, and then the inner product with the analysis signal $p(t)$ at different scales.

\begin{equation}
    WT_s(\epsilon,\omega)=\frac{1}{\sqrt{n}}\int_{- \infty}^{+\infty}p(t)\phi(\frac{t-\omega}{n})dt
\end{equation}
where $\epsilon$ is the scale factor (> 0) to stretch the each basic wavelet $\phi(t)$. $\omega$ is the displacement.
Mallat algorithm \cite{b23} provides an effective way to display DWT to process the data using the low-high-pass filters:

\begin{equation}
  	oL=\sum_{i=-\infty}^{\infty}T(i)\psi_l(2n-i)  
\end{equation}

\begin{equation}
    oH=\sum_{i=-\infty}^{\infty}T(i)\psi_h(2n-i)
\end{equation}
Where $T(i)$ means the signal. $\psi_l$, $\psi_h$, oL, oH are the low-pass filter, high-pass filter, output of low-pass filter, and output of high-pass filter, respectively. Notably, in the wavelet domain, the coefficient corresponding to the effective signal is large, and the coefficient corresponding to the noise is small. As a result, the noise can be removed by the threshold. In this paper, we apply the common $rigrsure$ threshold in DWT:
\begin{equation}
    g(k)=\big[sort|t| \big]^2,  \quad(k=0,1,...,N-1)
\end{equation}
In the equation, the absolute value of each signal is achieved and then sorted, and the square of each number is taken to obtain a new signal sequence.

\begin{equation}
    \gamma_t=\sqrt{g(k)}, \quad(t=0,1,...,{N-1})
\end{equation}

\begin{equation}
    Risk(t)=\frac{(N-2t+\sum_{i=1}^tg(j)+(N-t)f(N-t))}{N}
\end{equation}

\begin{equation}
    \gamma_t=\sqrt{g(t_{min})}
\end{equation}

The $t$ is the signal, $\gamma_t$ is the threshold and $Risk(t)$ is the generated risk. Take the minimum $g(t)$ corresponding to all risks $r(K)$ to get the final threshold $\gamma_t$.

\begin{figure}
    \centering
    \includegraphics[width=1.0\linewidth]{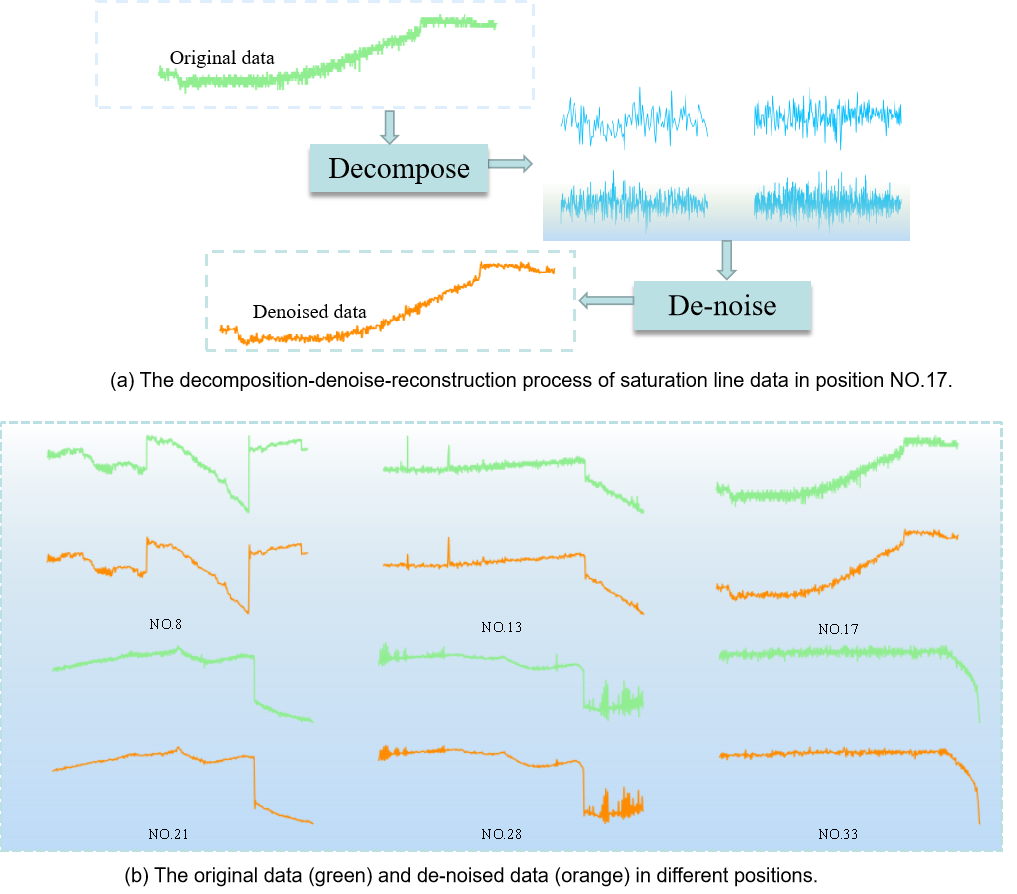}
    \caption{The decomposition-denoise-reconstruction process of saturation line data. It is obviously that after DWT process, the random noise of original data (green) is deleted and the data become smooth (orange).}
    \label{DWT_trans}
\end{figure}

The 3-level decomposition and the reconstruction process of DWT using Mallat algorithm is shown in Figure~\ref{DWT_trans}(a) and Figure~\ref{DWT_trans}(b), respectively. From Figure~\ref{DWT_trans}(a) we can see that after decomposing the signal into three different levels. In more detail, at the first level, the original signal $T$ is decomposed to the detail coefficients $oL_1$ and $oH_1$. Then the achieved $oL_1$ is decomposed to the other two coefficients $oH_1$ and $oL_1$ at the second level. The decomposition process does not end until the set number of $n$-level steps is reached. The Figure~\ref{DWT_trans}(b) illustrates the process of de-noise and reconstruction. The noises are shown with small wavelet coefficients, while the useful signals are shown with small wavelet coefficients. The time-series signal $T$ passes through the low-pass filter $oL_1$ and high-pass filter $oH_1$ for removing the wavelet coefficients of lower amplitude and restore the wavelet coefficients of higher amplitude to achieve the effect of noise reduction. Subsequently the wavelet reconstruction and integration process is applied on all of these coefficients. Employing the coefficient $oL_3$, the low frequency and high amplitude $rL_3$ is reconstructed. As shown in $rL_3$ in Figure~\ref{DWT_trans}, the sequences become smooth, showing the refined sequence patterns.  

\begin{figure}
    \centering
    \includegraphics[width=1.0\linewidth]{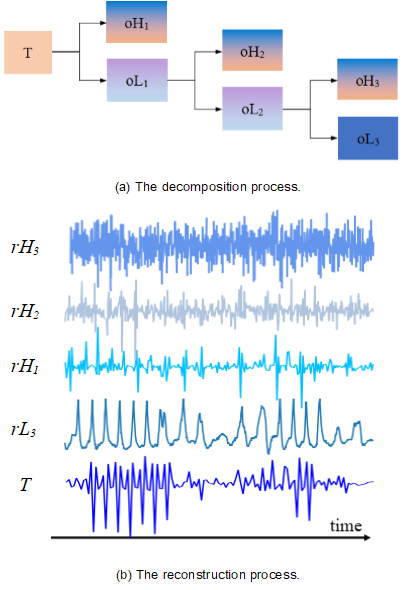}
    \caption{The Discrete Wavelet Transform process.}
    \label{fig:my_label}
\end{figure}

\begin{figure*}
    \caption{The channel and temporal attention. The first row indicates the channel-wise attention, while the second row indicates the temporal-wise attention. }
    \centering
    \includegraphics[scale=0.26]{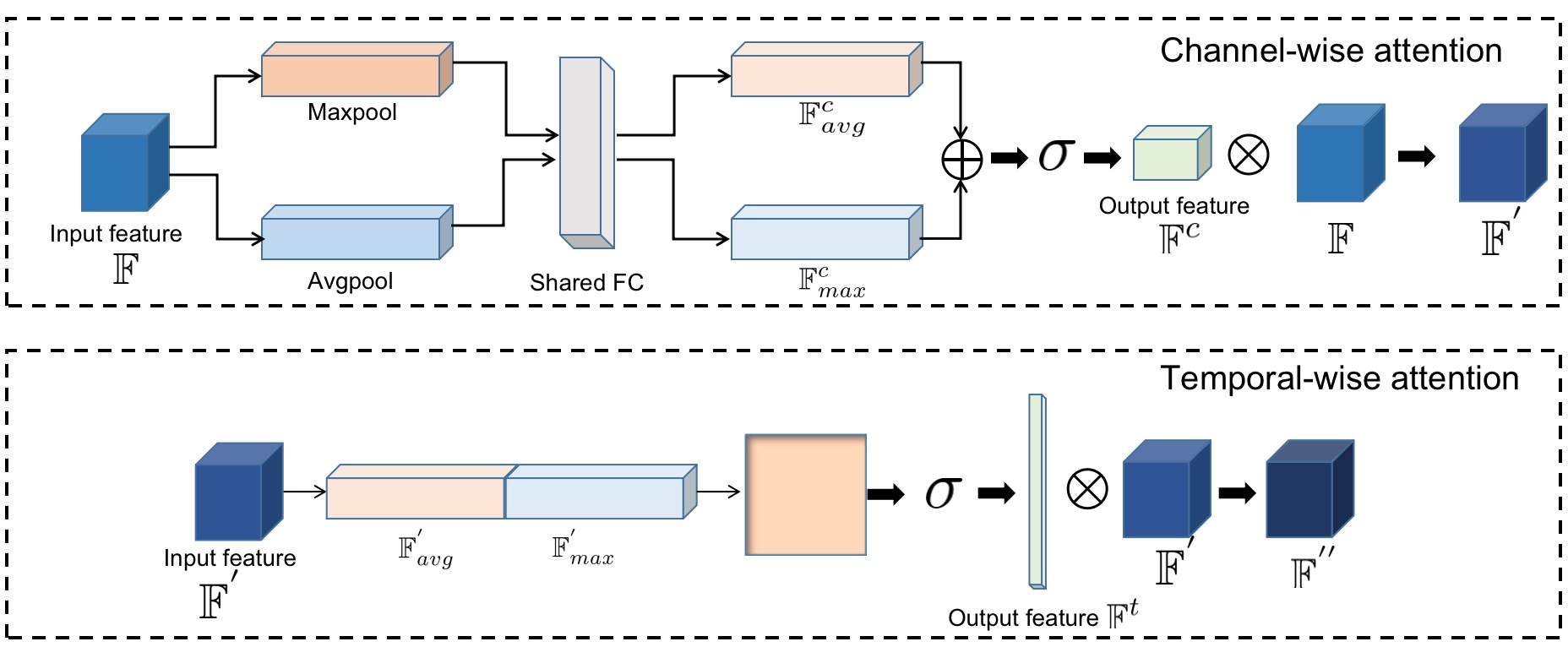}
    \label{channel-spatial}
\end{figure*}
\subsection{Structure of the proposed model}
\label{proposed}
Our study aims to develop the construction of a prediction system for forecasting the saturation line utilizing state-of-the-art LSTM and CNN networks. What has devoted to the popularity of the convolutional layer is the fact that it good at extracting and recognizing as well as identifying the structures of the time series in the monitoring data, while the LSTM networks achieve good performance in detecting long-short-term dependence. In light of this, the principle idea of our study is to combine the advantages of CNN and LSTM.\\
\indent We firstly bulid a baseline model namely CNN-LSTM, which utilize one CNN and one LSTM model. In further study, our proposed model improves the baseline with one more LSTM model together with two channel and temporal attention. As we mentioned before, our proposed model is a two-stage model, where the first part is the DWT process, the second part is the time-series prediction model. The convolutional layers encode the time-series information as high-dimensional structure, while the LSTM layer decodes the information from convolutional layers for time-series dependence. The channel and temporal-wise features are enhanced by the attention model. Our proposed model is shown in Figure~\ref{framework}.

Specifically, our proposed model includes one convolutional layer filters of 32, a max-pooling layer filters of 2, two LSTM layers of 25, 50, a flatten layer and a fully-connected layer in order. Different parameters of CNN and LSTM are compared for further study in Table.~\ref{hyper}. Furthermore, the attention module is embeded in the baseline CNN-LSTM model. The details of the embedded channel and temporal attention is described in Sec.~\ref{channel_temporal}.

\subsection{Channel and Temporal attention module}
\label{channel_temporal}
In this section, we explain the channel and temporal attention we utilized in this paper. The channel-wise operation is utilized for extracting the channel structure generated by the CNN model in high dimension. 

For the channel-wise features, the spatial information of a feature map is aggregated by the average-pooling and max-pooling operations, generating two different spatial descriptors: $\mathbb{F}^c_{avg}$ and $\mathbb{F}^c_{max}$, which represent average-pooled features and max-pooled features respectively.
\begin{equation}
    \begin{split}
        \mathbb{F}^c &= \sigma(\mathbf{FC}(AvgPool (\mathbb{F}) + \mathbf{FC}(Maxpool (\mathbb{F}) ) ) \\
        & = \sigma( \mathbf{W_1}(\mathbb{F}^c_{avg}) + \mathbf{W_1}(\mathbb{F}^c_{max}) ) ),
    \end{split}
\end{equation}

where $\sigma$ indicates the sigmoid function and $\mathbf{FC}$ indicates the shared fully connected layer with $\mathbf{Relu}$ function.

The final channel-wise attention is formulated as:
\begin{equation}
   \mathbb{F}^{'} = \mathbb{F}^{c} \otimes \mathbb{F},
\end{equation}
where $\otimes$ means element-wise product.

For the time-series prediction task, the temporal information is crucial since it expresses the implicit time series dependencies in a large degree. For the temporal-wise features, the channel information of a feature map by using two pooling operations are generated as two maps. 

\begin{equation}
    \begin{split}
        \mathbb{F}^t &= \sigma(\mathrm{Conv_{7 \times7}}(\mathrm{Concat}\big[ Avgpool (\mathbb{F}), Maxpool (\mathbb{F}) \big])) \\
        & = \sigma(\mathrm{Conv_{7 \times7}}(\mathrm{Concat} \big[ \mathbb{F}^t_{avg}, \mathbb{F}^t_{max} \big ])),
    \end{split}
\end{equation}
where $Conv_{7 \times7}$ represents the convolution kernel with the size of $7 \times7$.

The overall channel and temporal-wise attention is formulated as:
\begin{equation}
   \mathbb{F}^{''} = \mathbb{F}^{t} \otimes \mathbb{F{'}}.
\end{equation}

\section{Experiment and Results}
	% 	\subsection{Methods and Results of Machine Learning Models}
	% 	\indent By using RF algorithm, training and evaluation on the geologic hazard data are discussed in this section. In this part, we take M8, which is a feature of saturation line as an example.\\
	% 	\indent Metric in regression: The metric, coefficient of determination, denoted as ${R^2}$, is used in our evaluation methodology. It is the proportion of the total variation of the dependent. The  ${R^2}$ is defined as:
	% 	\begin{equation}
	% 	R^{2}=1-\frac{\sum_{i=0}^{m}(y_i-\widehat{y_i})^2}{\sum_{i=0}^{m}(y_i-\overline{y_i})^2}
	% 	\end{equation}
	% 	Where $y_i$ represents the true value, $\widehat{y_i}$ represents predictive value and $\overline{y_i}$ represents average of true values. The better the model performance comes the higher the ${R^2}$.

\subsection{Metrics}
\begin{figure*}
    \centering
    \begin{minipage}[c]{0.33\linewidth}
        \centering
        \centerline{
        \includegraphics[width=1.06\textwidth]{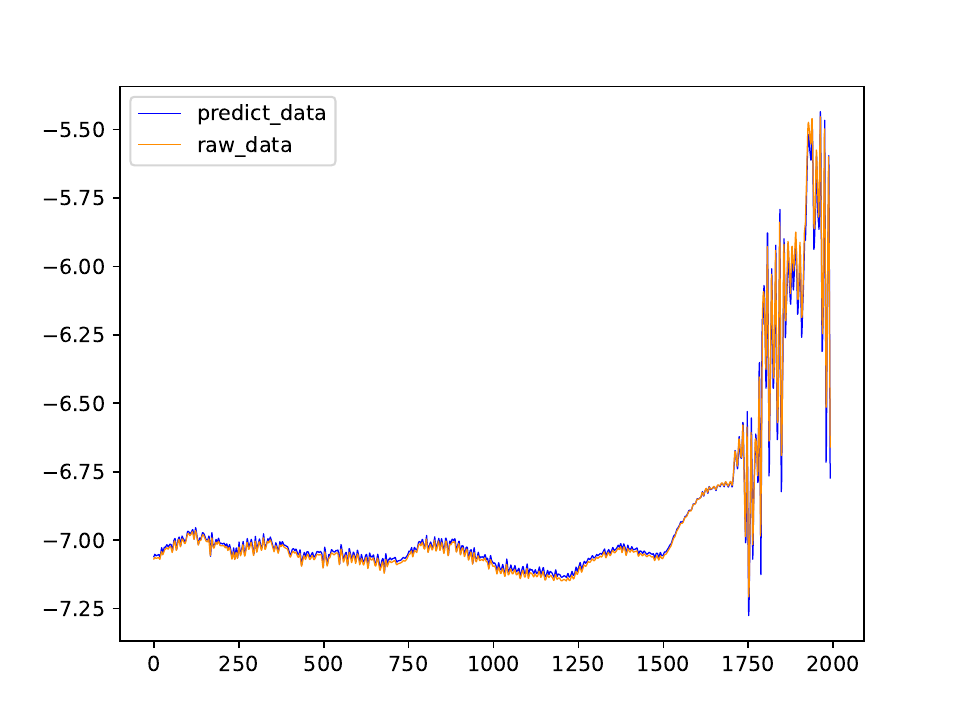}
        }
        \centerline{(a) No.8}
    \end{minipage}\vspace{5pt}
    \begin{minipage}[c]{0.33\linewidth}
        \centering
        \centerline{
        \includegraphics[width=1.06\textwidth]{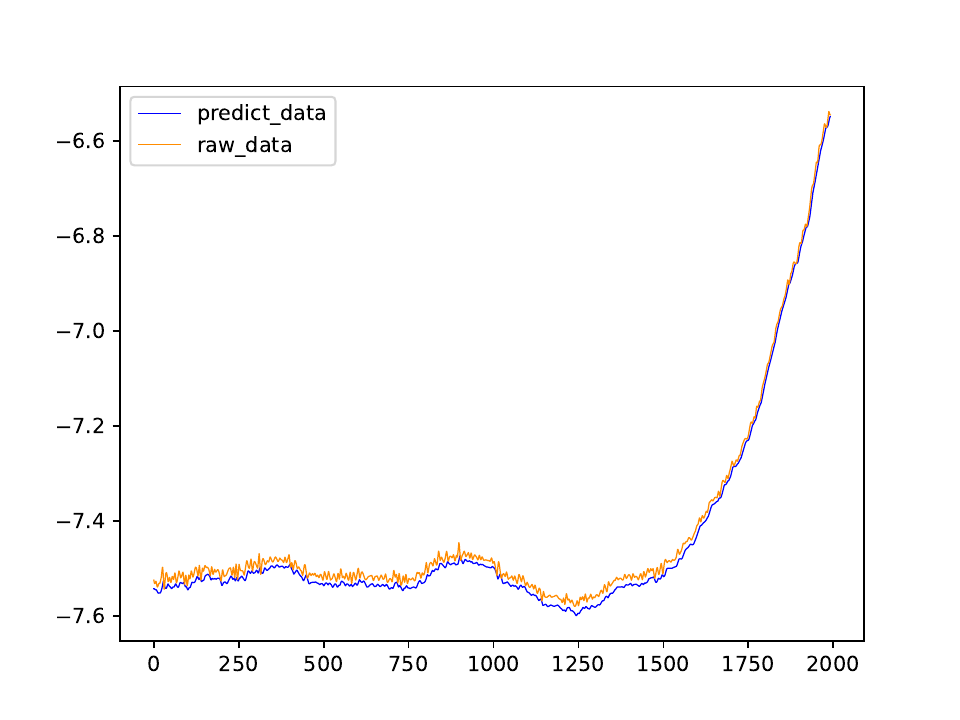}
        }
        \centerline{(b) No.13}
    \end{minipage}
    \begin{minipage}[c]{0.33\linewidth}
        \centering
        \centerline{
        \includegraphics[width=1.06\textwidth]{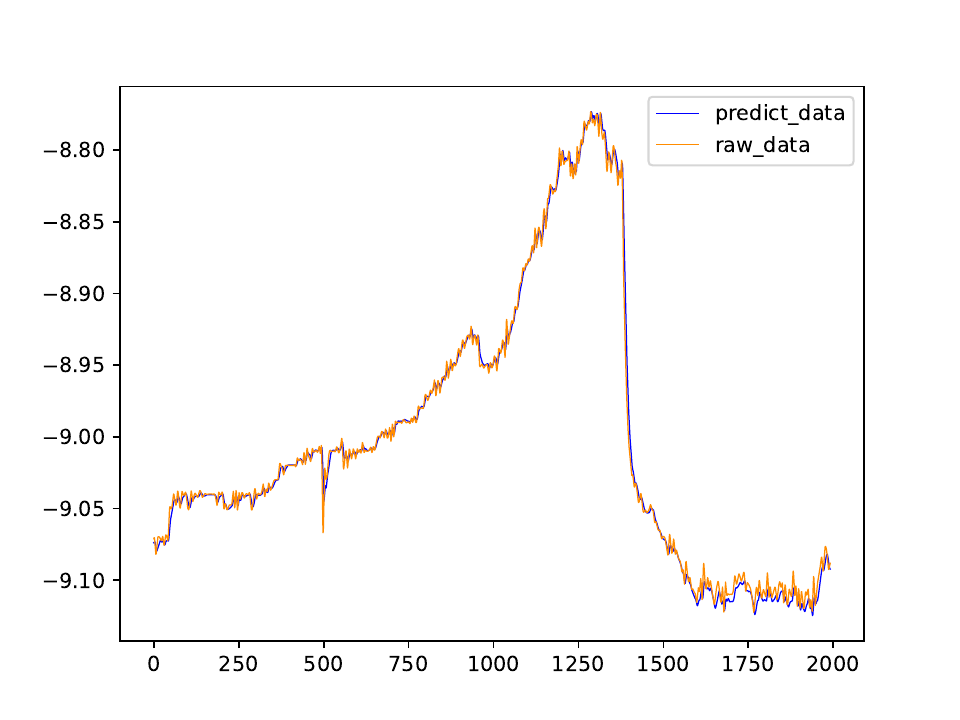}
        }
        \centerline{(c) No.17}
    \end{minipage} 
    
    \begin{minipage}[c]{0.33\linewidth}
        \centering
        \centerline{
        \includegraphics[width=1.06\textwidth]{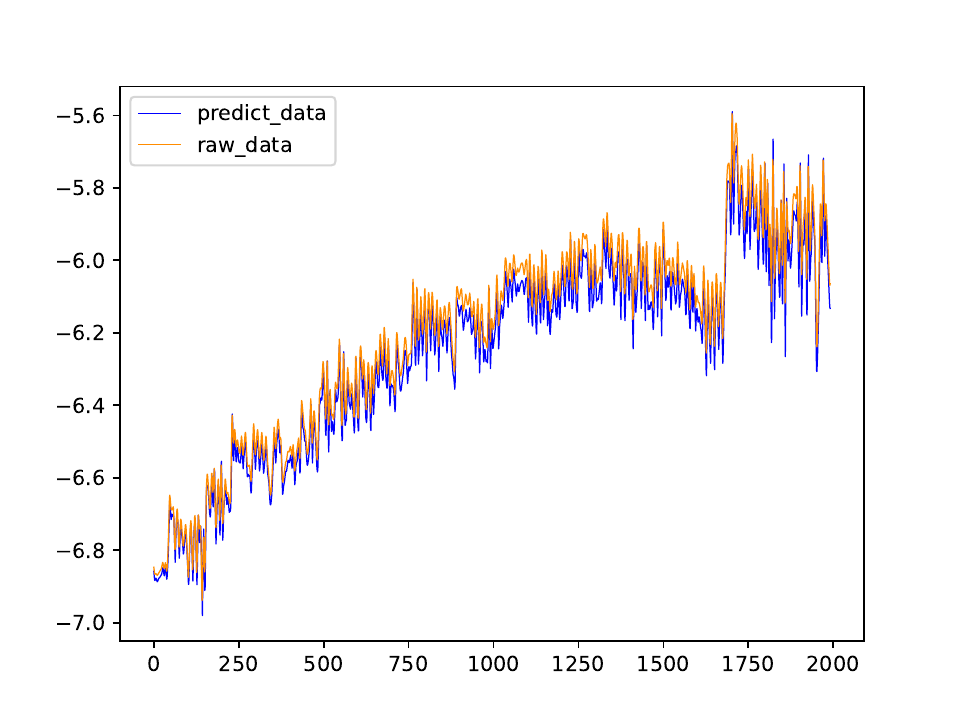}
        }
        \centerline{(d) No.21}
    \end{minipage} 
     \begin{minipage}[c]{0.33\linewidth}
        \centering
        \centerline{
        \includegraphics[width=1.06\textwidth]{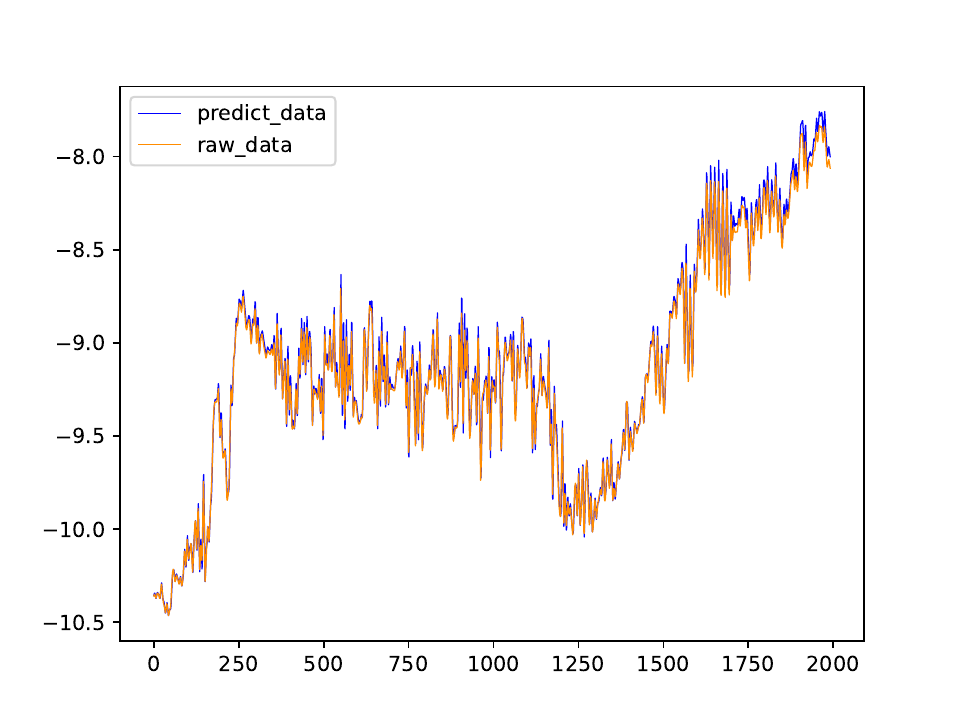}
        }
        \centerline{(e) No.28}
    \end{minipage} 
     \begin{minipage}[c]{0.33\linewidth}
        \centering
        \centerline{
        \includegraphics[width=1.06\textwidth]{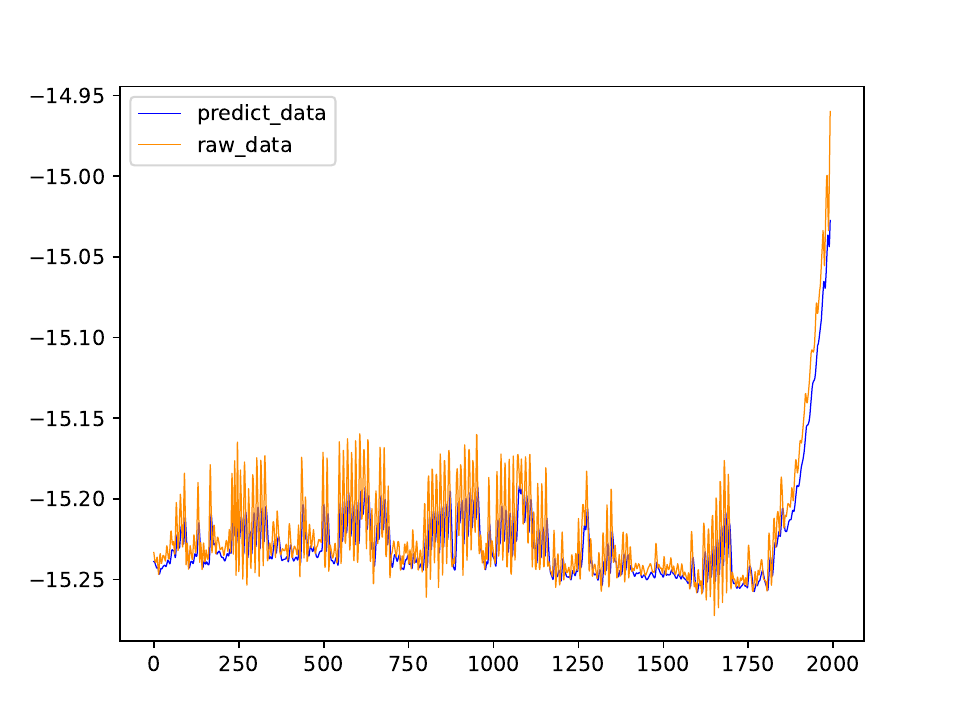}
        }
        \centerline{(f) No.33}
    \end{minipage} 
     \caption{The prediction scatters of saturation line at different positions using our proposed model.}
\label{prediction_result}
\end{figure*}

The prediction performance of our proposed model is evaluated by root mean square error (RMSE), mean absolute percentage error (MAPE) and coefficient of determination (R${^2}$). In fact, RMSE meets an important problem: considering that although the model has an error of less than 0.1\% in the 95\% dataset and very big error in the other 5\% dataset, the overall RMSE is still high, resulting in this model considered as a poor model. To solve this problem, MAPE is utilized in the evaluation process. It is the proportion of the total variation of the dependent.
	\begin{equation}
	\mathrm{RMSE} = \sqrt{\frac{1}{n}\sum_{i=1}^n(y_t-\hat{y_p})^2}
	\end{equation}
	
	\begin{equation}
	\mathrm{MAPE}=\sum_{i=1}^n\frac{1}{n}|\frac{y_t-\hat{y_p}}{y_t}|*100\%
	\end{equation}
	
	\begin{equation}
		\mathrm{R^{2}}=1-\frac{\sum_{i=0}^{n}(y_t-\hat{y_p})^2}{\sum_{i=0}^{n}(y_t-\overline{y_t})^2}
		\end{equation}
	Where $y_t$ represents the true value, $\hat{y_p}$ represents predicted saturation line value, $\overline{y_t}$ represents average of true value, and n is the count of data. Figure.~\ref{prediction_result} shows the prediction results of our proposed model on five different real-world dataset.
	
\subsection{Implement details}
We implement our network using the Tensorflow deep learning framework, and all the models are trained with two Nvidia Tesla V100 GPUs. In this study, the proposed model is trained for 200 epochs with a batch size of 32. We apply RMSE as loss function and Adam for optimizer. The Adam is an improved RMSProp optimizer combining with the moments trick. It is worth noticing that in order to reduce the feature loss during the convolutional layers, same padding operation was conducted during this process. 
Specifically, we set the sequence length as 10. On the one hand, considering that a longer sequence length will occupy a huge computer memory, on the other hand, we found through experiments that set the sequence length to 10 achieves better performance than, $e.g.$, 3, 7, 10, 20, 50. The most important thing for the hyperparameter selection of the model is the learning rate of the network, which has a
significant influence on time consumption until convergence. If the learning rate is set too large, the loss function will be difficult to converge, resulting in a lower final detection accuracy; On the contrary, a small learning rate will lead to slow convergence and increase the training time. For the optimizer, we use the Adam with a momentum of 0.9 and a weight decay of 0.001. For the selection of the number of network iterations, the training process is stopped when the model no longer converges.

\section{Experiment}

\begin{figure*}
    \centering
    \begin{minipage}[c]{0.3\linewidth}
        \centering
        \centerline{
        \includegraphics[width=1.06\textwidth]{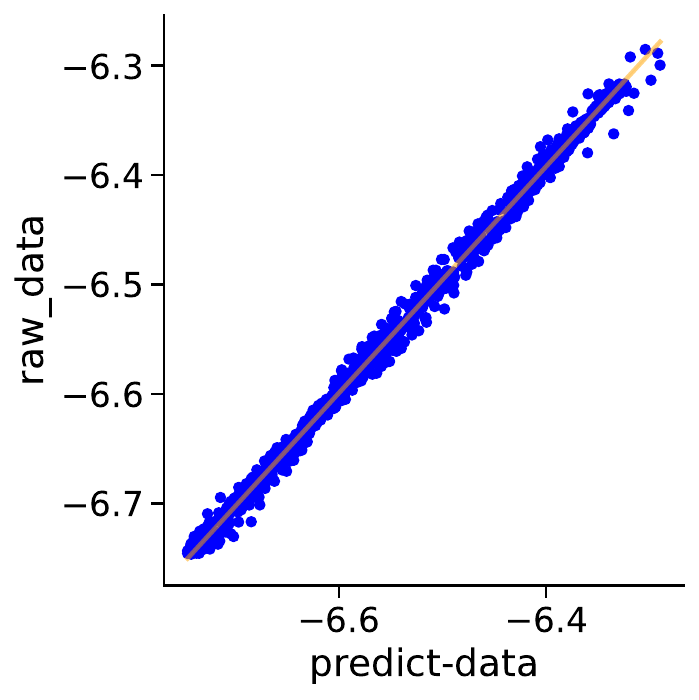}
        }
        \centerline{(a) No.8}
    \end{minipage}\vspace{5pt}
    \begin{minipage}[c]{0.3\linewidth}
        \centering
        \centerline{
        \includegraphics[width=1.06\textwidth]{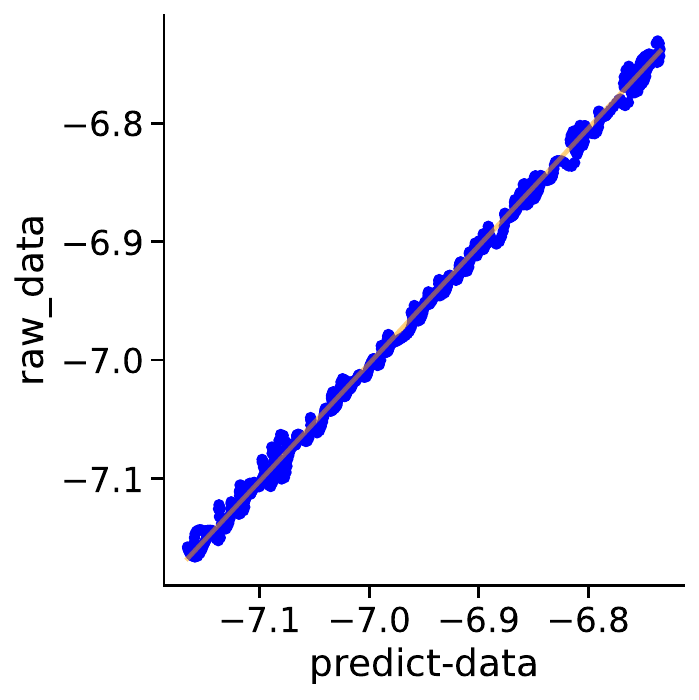}
        }
        \centerline{(b) No.13}
    \end{minipage}
    \begin{minipage}[c]{0.3\linewidth}
        \centering
        \centerline{
        \includegraphics[width=1.06\textwidth]{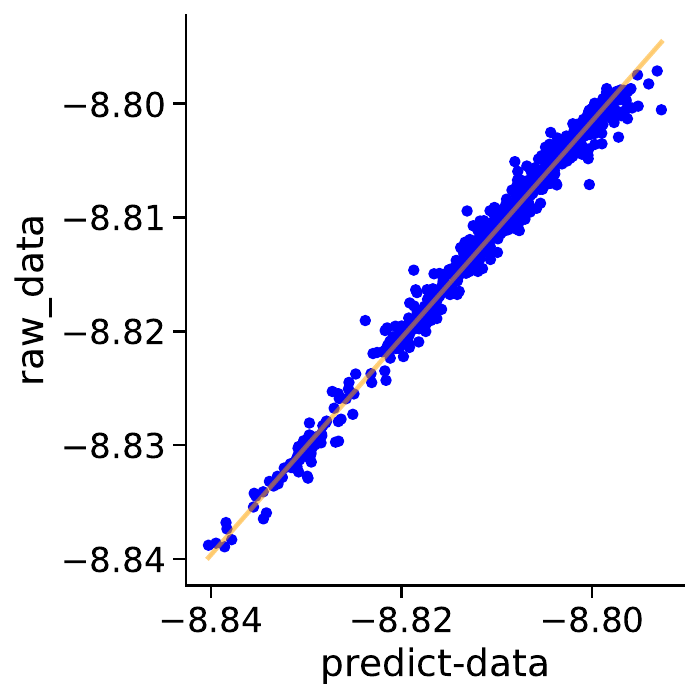}
        }
        \centerline{(c) No.17}
    \end{minipage} 
    
    \begin{minipage}[c]{0.3\linewidth}
        \centering
        \centerline{
        \includegraphics[width=1.06\textwidth]{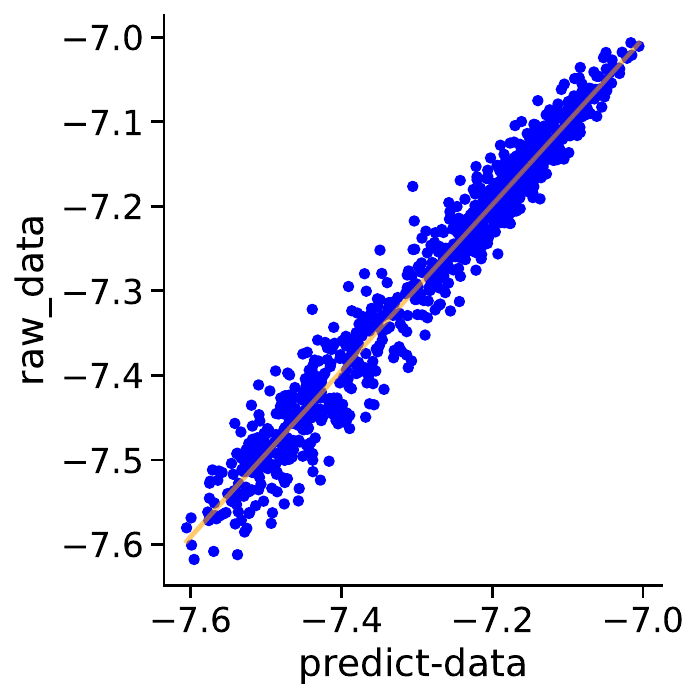}
        }
        \centerline{(d) No.21}
    \end{minipage} 
     \begin{minipage}[c]{0.3\linewidth}
        \centering
        \centerline{
        \includegraphics[width=1.06\textwidth]{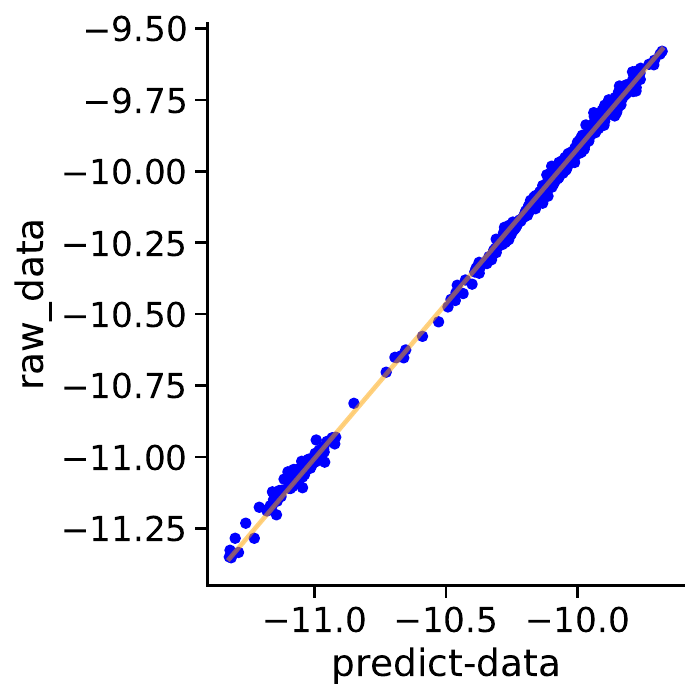}
        }
        \centerline{(e) No.28}
    \end{minipage} 
     \begin{minipage}[c]{0.3\linewidth}
        \centering
        \centerline{
        \includegraphics[width=1.06\textwidth]{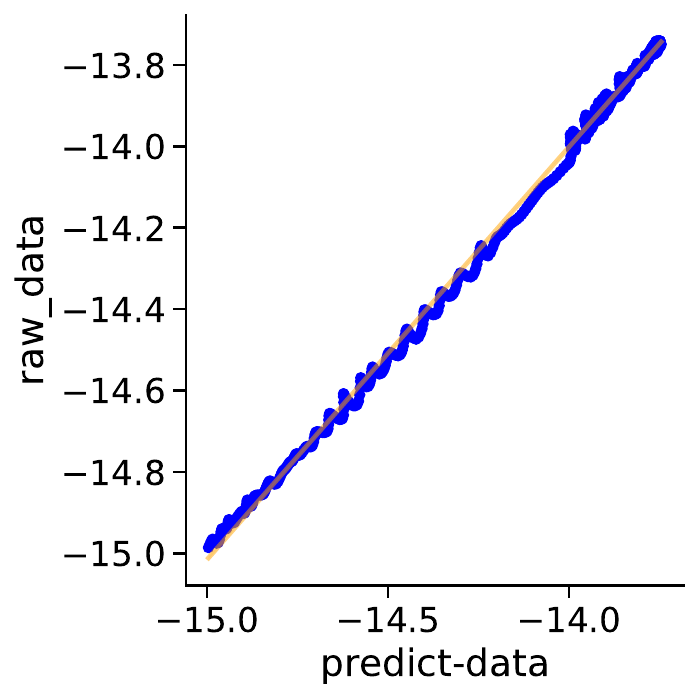}
        }
        \centerline{(f) No.3}
    \end{minipage} 
     \caption{The prediction scatters of saturation line at different positions using our proposed channel and temporal attention model.}
    \label{improved}
\end{figure*}

Our proposed model is evaluated and compared to other models to show the prediction performance. The prediction performance of our proposed is shown in Table.~\ref{test_result}, where NO.8, NO.13, NO.17, NO.21, NO.28, NO.33 mean the different station of saturation line mentioned above. The deep and abstract features the convolutional layer learned may be different from the ordinary time-series information from the raw data. This is obviously a disadvantage when the monitoring data contains only simple or even linear information. While using one convolutional layer and two LSTM layers together with the channel and temporal attention can capture the long-short-term data dependencies to a significant degree from the result.
The scatter plots of raw data and predicted
saturation line is illustrated in Figure.~\ref{improved}, which helps show the prediction performance intuitively.

\begin{table}
\caption{Prediction performance of the proposed model on MAPE, RMSE and R$^2$.}
\centering
\begin{tabular}{ccccccc}
\toprule \hline
\textbf{Metrics}	& \textbf{NO.8}& \textbf{NO.13} &\textbf{NO.17} &\textbf{NO.21} &\textbf{NO.28} &\textbf{NO.33} \\ \hline
	RMSE   &0.0209 &0.030 &0.0336 &0.0170 &0.0123 & 0.0366              \\
	MAPE    &3.346 &3.316 &3.207 &1.589 &3.221 &3.432  \\
	\textbf{R$^2$}     &\textbf{0.969} &\textbf{0.974} &\textbf{0.937} &\textbf{0.981} &\textbf{0.951} &\textbf{0.892} \\ 
\bottomrule
\label{test_result}
\end{tabular}
\end{table}

\begin{table}
	\caption{Performance comparison of different machine learning and deep learning models. For a fair comparison, we only compare the runtime of deep models.}
	\centering
	\resizebox{1\linewidth}{!}{   
	\begin{tabular}{ccccccc}
	\toprule \hline
	\textbf{Type} & \textbf{RMSE}& \textbf{MAPE} &\textbf{R$^2$} &\textbf{Runtime (s)} &\textbf{Model}\\ 
	\midrule
	
	\multirow{3}{*}{Shallow model} &0.132 &4.542 &0.548 & $-$ &SVR\\
	&0.141 &4.312 &0.489 & $-$ &DTR\\
	&0.251 &4.186 &0.839 & $-$ &RFR\\

	\multirow{5}{*}{Deep model} &0.0504 &3.744 &0.798 &44.08 &MLP\\
	&0.0308 &3.645 &0.864 &47.54 &RNN\\
	&0.0221 &3.602 &0.879 &63.55 &GRU  \\
	&0.0214 &3.596 &0.887  &77.08 &LSTM\\ 
	&\textbf{0.0209}  &\textbf{3.346}  &\textbf{0.969} &\textbf{25.49} &\textbf{Ours}\\  \hline
		\bottomrule
		\label{compare}
		\end{tabular}}
	\end{table}

\begin{table*}
		\caption{Prediction cases using different hyperparameters in our proposed model. }
		\centering
		%% \tablesize{} %% You can specify the fontsize here, e.g., \tablesize{\footnotesize}.
		\begin{tabular}{cccccccccc}
		\toprule \hline
		\ & \textbf{$ $}	& \textbf{Batch Size}& \textbf{Convolution kernel Size} &\textbf{Pooling Size} &\textbf{LSTM Units} &\textbf{RMSE} &\textbf{MAPE} &\textbf{R$^2$} &\textbf{Runtime (s)}\\
		\midrule
	&Case 1 &32 &16 &2 &[25,50] &0.0411  & 3.352  &0.899  &50.87   \\
	&Case 2 &64 &32 &2 &[50,75] & 0.0208 &3.324 &0.971 &49.37 \\
	&Case 3 &32 &16 &4 &[25,50] &0.0514  &4.114  &0.792 &21.13  \\ 
	&Case 4 &64 &32 &4 &[50,75] &0.0504  &4.011   &0.801 &21.88 \\
	&Case 5 &16 &16 &2 &[50,50] &0.0296  &3.322 &0.972 &51.32 \\
	&Case 6 &128 &16 &4 &[25,50] &0.0521  &4.281   &0.784 &25.12 \\
	&Case 7 &16 &32 &2 &[25,75] &0.0385  &3.513   &0.902  &50.52\\
	&Case 8 &128 &32 &2 &[25,50] &0.0311  &3.501  &0.932 &37.49\\
	&Case 9 &64 &32 &2 &[25,50] &0.0209  &3.346  &0.969& 25.49\\  \hline
		\bottomrule
		\label{hyper}
		\end{tabular}
\end{table*}

To show the superiority of our proposed model, we apply comparative studies with other state-of-the-art machine learning and deep learning models, including the support vector regression (SVR), decision tree regression (DTR), random forest regression (RFR), multilayer perception (MLP), single GRU, simpleRNN as well as LSTM models on NO.8 dataset. Table.~\ref{compare} presents the RMSE, MAPE and R$^2$ score of these models in our experiments, which demonstrates that our proposed model significantly outperforms the others in R$^2$.

In order to build the complete saturation line prediction model and show the reliability of our proposed model together with parameters set, we compared different hyperparameters such as batch size, filters in of the convolutional layers, max-pooling size, number of LSTM cells in our experiments. Table.~\ref{hyper} lists the different situations of combing multiple hyperparameters. Considering several evaluation metrics and running time, we choose the design shown in Case 9. Specifically, in term of the evaluation metrics used in this task, although Case 2 and Case 5 achieve slight better performance than that in Case 9, the Runtime is almost twice. Predictions will perform worse in real time when the running time is excessive, especially when there is a large amount of data. The disadvantage is more pronounced for a large amount of data, and this incurs no loss of generality. Case 3 needs the least Runtime but achieve low accuracy. To be clear, according to Case 9, the implement is: the batch size is 64, one convolutional layer filters of 32, a max-pooling layer filters of 2, two LSTM layers of 25 and 50.

\begin{figure*}
    \centering
    \begin{minipage}[c]{0.3\linewidth}
        \centering
        \centerline{
        \includegraphics[width=1.06\textwidth]{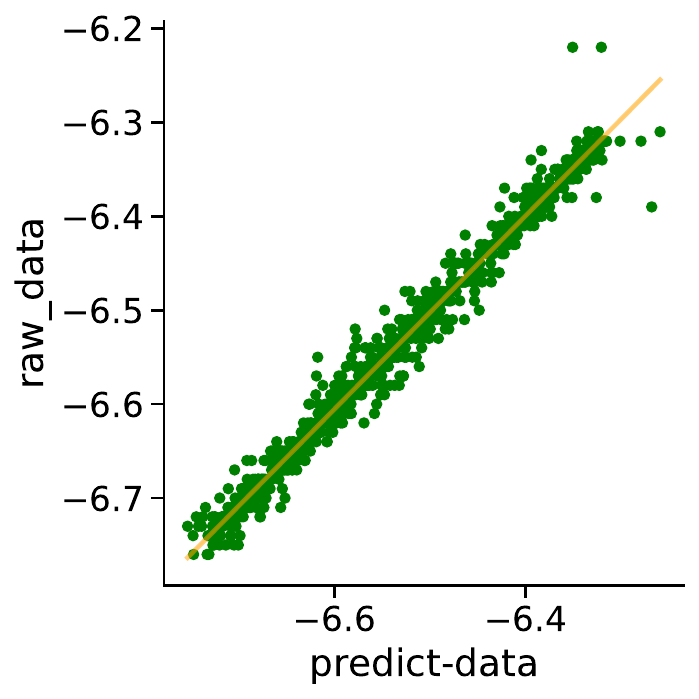}
        }
        \centerline{(a) No.8}
    \end{minipage}\vspace{5pt}
    \begin{minipage}[c]{0.3\linewidth}
        \centering
        \centerline{
        \includegraphics[width=1.06\textwidth]{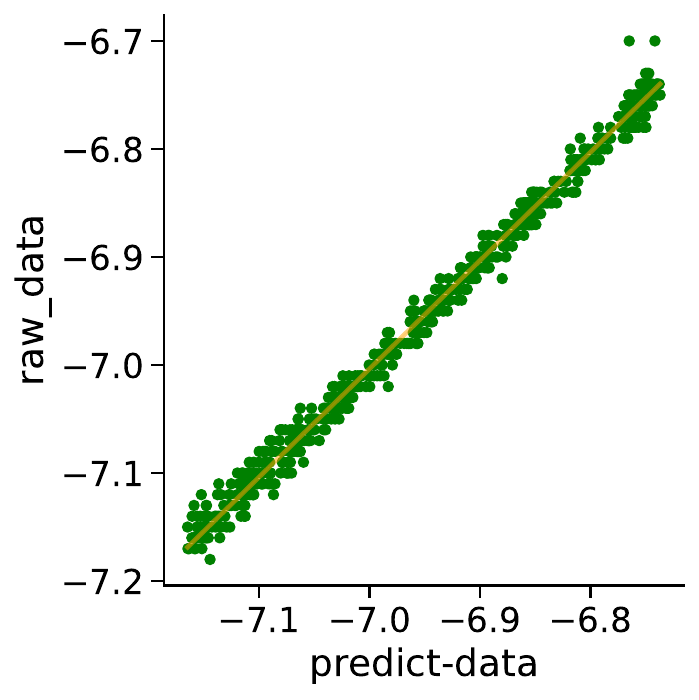}
        }
        \centerline{(b) No.13}
    \end{minipage}
    \begin{minipage}[c]{0.3\linewidth}
        \centering
        \centerline{
        \includegraphics[width=1.06\textwidth]{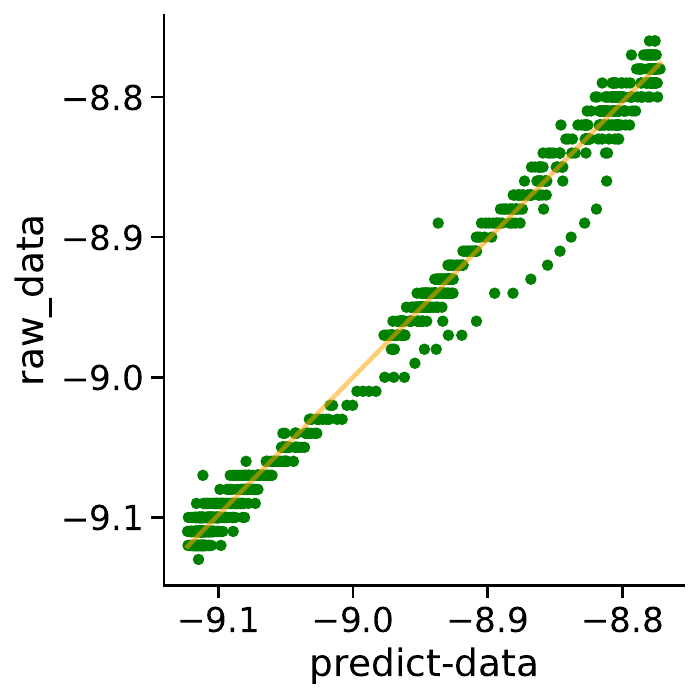}
        }
        \centerline{(c) No.17}
    \end{minipage} 
    
    \begin{minipage}[c]{0.3\linewidth}
        \centering
        \centerline{
        \includegraphics[width=1.06\textwidth]{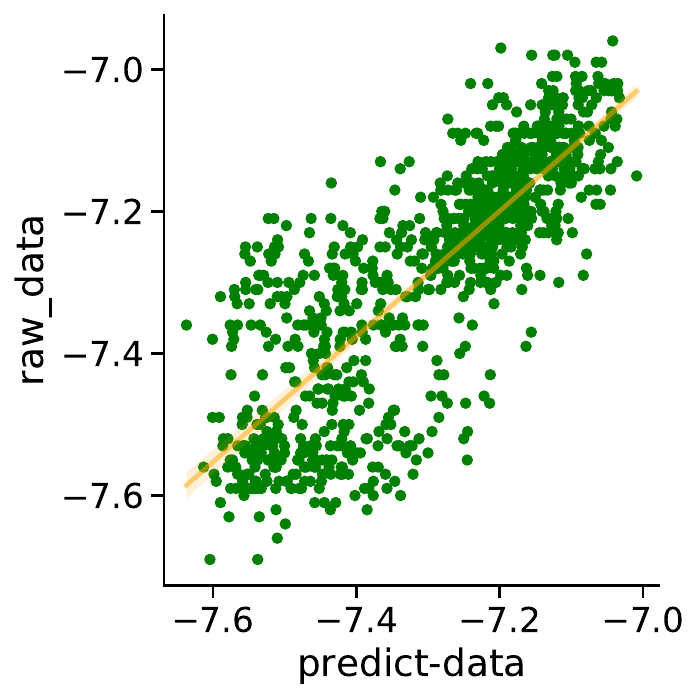}
        }
        \centerline{(d) No.21}
    \end{minipage} 
     \begin{minipage}[c]{0.3\linewidth}
        \centering
        \centerline{
        \includegraphics[width=1.06\textwidth]{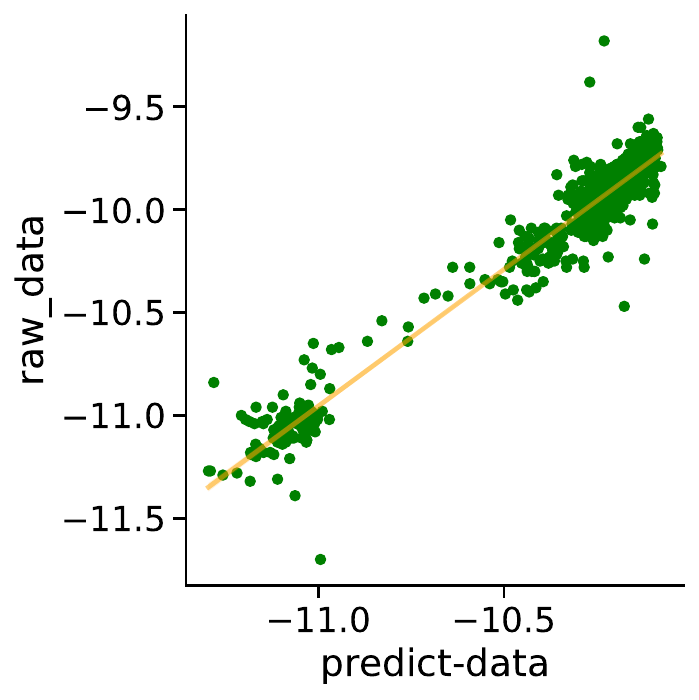}
        }
        \centerline{(e) No.28}
    \end{minipage} 
     \begin{minipage}[c]{0.3\linewidth}
        \centering
        \centerline{
        \includegraphics[width=1.06\textwidth]{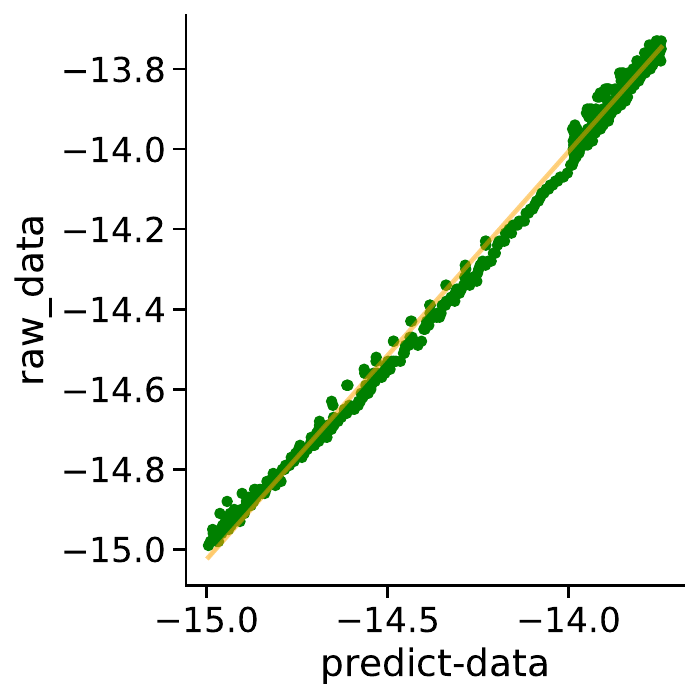}
        }
        \centerline{(f) No.33}
    \end{minipage} 
     \caption{The prediction scatters of saturation line at different positions utilizing the baseline model (CNN-LSTM).}
\label{original}
\end{figure*}

\section{ablation study}
We conduct the ablation study to evaluate the effectiveness of DWT and channel and temporal-wise attention. Since our proposed model includes the DWT process, one convolutional layer, two LSTM layers and two channel and temporal attention modules, we evaluate these experiments by removing these modules. The results display in Table.~\ref{ablation}.

Furthermore, we compare the fitting results with the beseline model (one CNN and one LSTM). It can be seen that our proposed model (Figure.~\ref{improved}) achieves more effective predictions than that utilizing baseline model (Figure.~\ref{original}), especially in the predictions NO.17, NO.21, and NO.28. This is because although the long-term and short-term dependence and hidden time series information can be discovered from the data, the prediction accuracy is greatly affected due to the presence of noise in the data. Furthermore, the channel and temporal-wise features are ignored by the simple CNN-LSTM model.

\begin{table*}[]
\caption{Ablation study. `-attention', `-DWT' and `-LSTM' mean remove the component. Baseline means the simple CNN-LSTM (one CNN and one LSTM) model.}
\begin{tabular}{ccccclcccccclc}
\toprule \hline 
\multicolumn{1}{l}{}   & Metrics              & -attention & -DWT  & -LSTM  & Baseline & Ours  &                        & Metrics              & -attention & -DWT  & -LSTM & Baseline & Ours  \\ \hline 
\multirow{3}{*}{No.8}  & RMSE                 & 0.038      & 0.032 & 0.027  & 0.045    & 0.021 & \multirow{3}{*}{No.21} & RMSE                 & 0.019      & 0.019 & 0.018 & 0.038    & 0.017 \\
                       & R$^2$  & 0.953      & 0.942 & 0.961  & 0.931    & 0.969 &     & R$^2$ & 0.957 & 0.959 & 0.951 & 0.851   & 0.961 \\
                       & MAPE                 & 3.658      & 3.719 & 3.813  & 4.132    & 3.346 &                        & MAPE                 & 1.698      & 1.688 & 1.649 & 2.051    & 1.589 \\ \hline
\multirow{3}{*}{No.13} & RMSE                 & 0.041      & 0.039 & 0.034  & 0.051    & 0.030 & \multirow{3}{*}{No.28} & RMSE                 & 0.021      & 0.024 & 0.019 & 0.029    & 0.012 \\
                       & R$^2$ & 0.962      & 0.967 & 0.968  & 0.951    & 0.974 &                        & R$^2$ & 0.937      & 0.932 & 0.941 & 0.928    & 0.951 \\
                       & MAPE                 & 3.517      & 3.481 & 3.399  & 3.698    & 3.316 &                        & MAPE                 & 3.412      & 3.521 & 3.309 & 3.712    & 3.221 \\ \hline
\multirow{3}{*}{No.17} & RMSE                 & 0.041      & 0.043 & 0.0378 & 0.061    & 0.034 & \multirow{3}{*}{No.33} & RMSE                 & 0.042      & 0.045 & 0.041 & 0.051    & 0.037 \\
                       & R$^2$ & 0.928      & 0.931 & 0.935  & 0.892    & 0.937 &                        & R$^2$ & 0.881      & 0.873 & 0.885 & 0.869    & 0.892 \\
                       & MAPE                 & 3.471      & 3.391 & 3.298  & 3.961    & 3.207 &                        & MAPE                 & 3.613      & 3.772 & 3.301 & 3.102    & 3.432 \\ \bottomrule
\end{tabular}
\label{ablation}
\end{table*}

\indent To overcome the drawbacks of baseline model that cannot de-noise the raw data, we applied the DWT to decompose the saturation line into different time-frequency sequences and remove the random noise. Subsequently, the data after noise reduction is trained by our proposed model.
With the help of enhancing the channel and temporal features, the results of all positions are shown in Table.~\ref{ablation}. This once again proves that the DWT method can remove a large amount of useless information, thereby assisting our model to more accurately explore the time series information hidden between the data. It also can be illustrated from Table.~\ref{ablation} and the comparison of Figure.~\ref{improved} and Figure.~\ref{original} that the our proposed model achieves the better performance than the baseline model at all saturation line stations.

For example, on dataset No.21, the baseline achieves the R$^2$ of 0.851, while our proposed model has the value of 0.961. When removing the DWT, the model has the value of 0.959 in terms of R$^2$; when removing the attention module, the model has the value of 0.957 in terms of R$^2$; when removing a LSTM, the R$^2$ drops form 0.961 to 0.951. 
The results shows that all the components are crucial for our proposed model. 

% \renewcommand\arraystretch{1.5}
% 	\begin{table}
% 	\caption{Performance of our proposed model on five datasets.}
% 	\centering
% 	\begin{tabular}{cccccccc}

% \hline \toprule
% Model Type                      & \textbf{Dataset} & \textbf{RMSE} & \textbf{MAPE} & \textbf{R$^2$} \\ \midrule
% \multirow{6}{*}{Proposed Model} & NO.8                              & 0.0205                         & 2.321                          & 0.988                           \\
%                                 & NO.13                             & 0.0231                         & 2.425                          & 0.984                           \\
%                                 & NO.17                             & 0.0046                         & 2.032                          & 0.986                           \\
%                                 & NO.21                             & 0.0161                         & 2.056                          & 0.995                           \\
%                                 & NO.28                             & 0.0044                         & 2.023                          & 0.996                           \\
%                                 & NO.33                             & 0.0039                         & 3.512                          & 0.968                           \\ \hline \bottomrule

% 	\end{tabular}
% 	\end{table}

\section{Discussion and Conclusion}

In this work, we applied a new method to predict the safety of tailings pond according to the saturation line using our proposed model, which is also first used in tailings pond risk prediction. Compared with the traditional methods, the risk evaluation method of tailings ponds has the characteristics of high accuracy and high real-time performance. 

The contributions of this work is three fold: First, proposing an effective channel and temporal attention-based CNN-LSTM network to predict the saturation line, which achieves satisfied performance in terms of MAPE, RMSE and R$^2$. Second, comparing our proposed model with different hyperparameters and with other state-of-the-art models. Third, conducting the ablation studies to confirm the components of our model. The wavelet transform method is applied to overcome the shortcomings that the original CNN-LSTM model could not de-noise the data. The wavelet transform decomposes the data into 4 layers of wavelets, selects the $rigrsure$ threshold to de-noise the decomposed wavelets and then reconstructs them, subsequently feeds the reconstructed useful signals to our attention-based model to obtain better prediction results.

In tailings pond risk prediction task, these experiments consequently provide applicability of the proposed model. Additionally, our model can also be used to predict water levels, weather, and air quality as time-series predictions. The model is evidently capable of not only extracting and recognizing spatial and time series structures, but also identifying long-term and short-term series information.

\indent Our method applies one factor, and in the future, we will focus on more factors of the safety monitoring parameters of the tailings pond, such as the underground displacement, ground displacement and dry beach length. We should also build a risk level that corresponds to the tailings pond monitoring so that our future work more intuitively reflects the safety of the tailings pond.

		\vspace{6pt} 
		
		
		%=====================================
		% References, variant A: internal bibliography
		%=====================================

\bibliographystyle{IEEEtran}      %LaTex Class文件, IEEEtran为给定模板格式定义文件名

\begin{IEEEbiography}[{\includegraphics[width=1in,height=1.25in,clip,keepaspectratio]{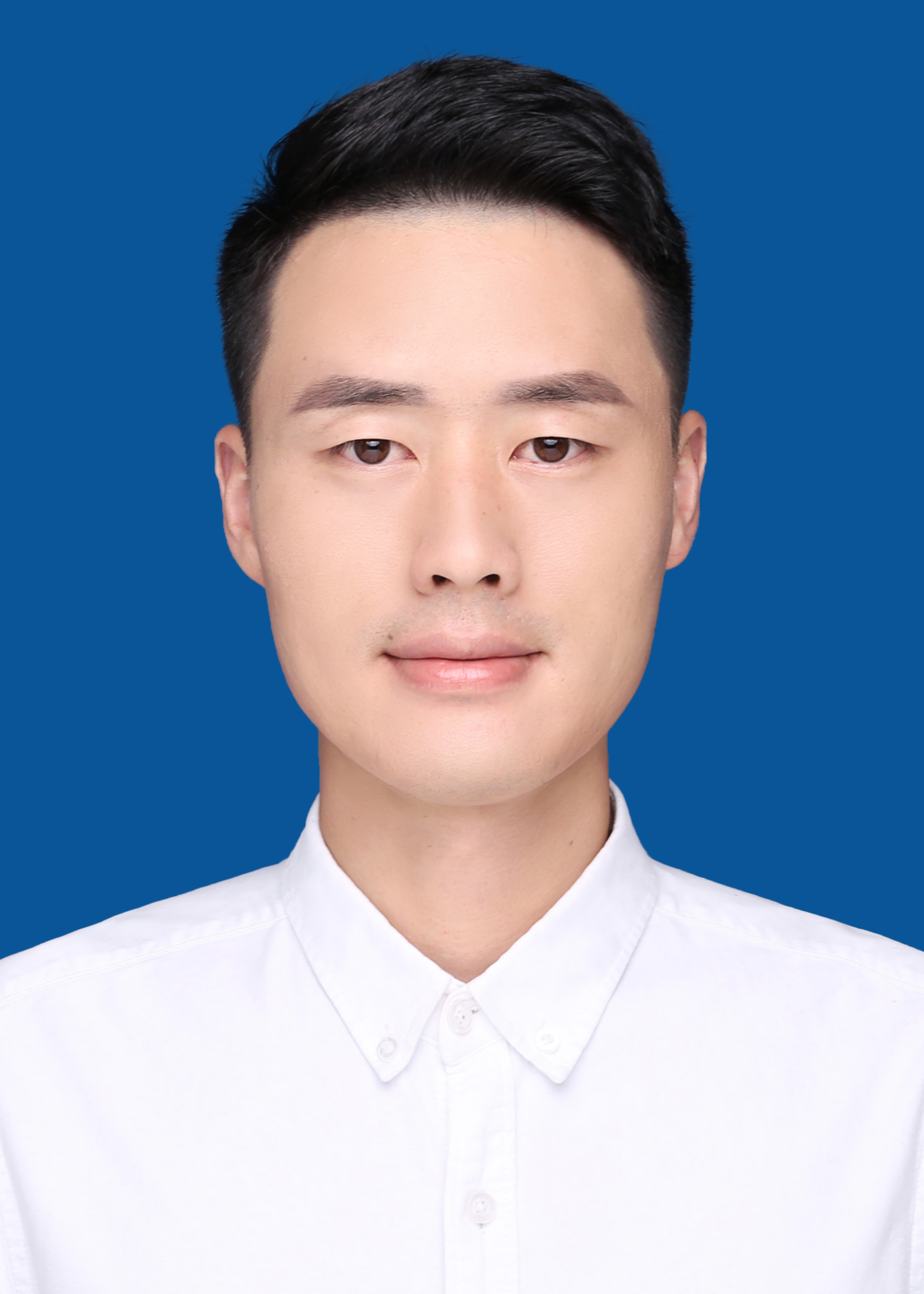}}]WEI WANG received the M.S. degree from the School of Mechanical and Electrical Engineering, China Jiliang University. He is currently an Algorithm Engineer at Zhejiang PeckerAI Technology., Ltd. His research interests include machine learning, image processing, Software development.
\end{IEEEbiography}
\begin{IEEEbiography}[{\includegraphics[width=1in,height=1.25in,clip,keepaspectratio]{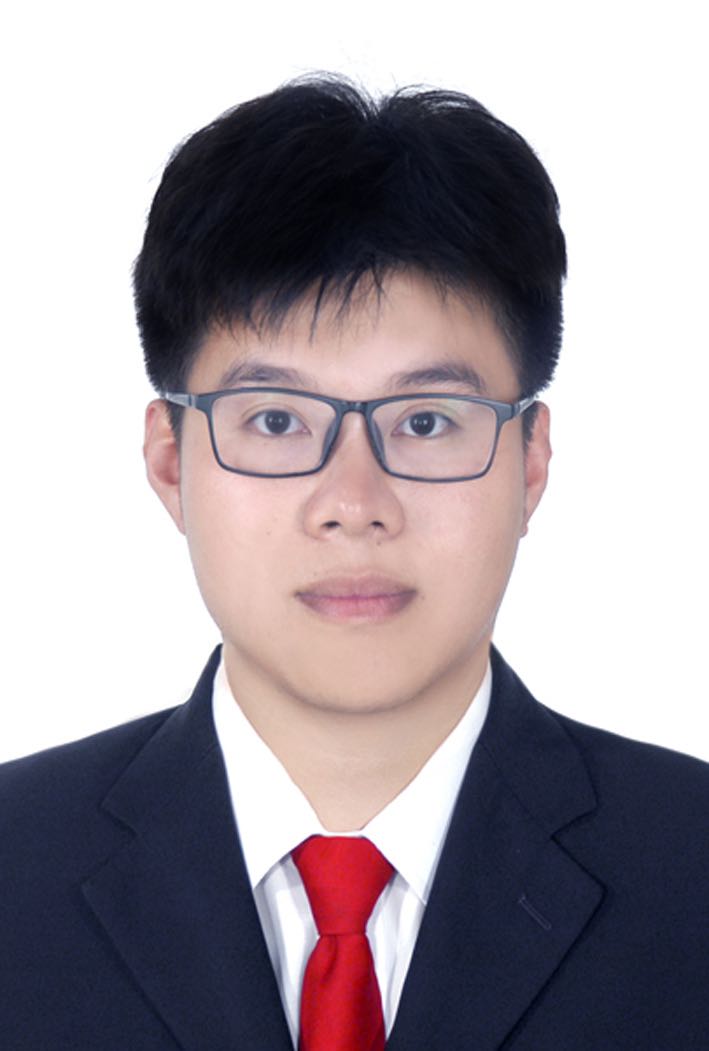}}]LINCHAO LI received the M.S degree China Jiliang University. He is a machine learning algorithm engineer. His research interests include computer vision, image processing.
\end{IEEEbiography}

\begin{IEEEbiography}[{\includegraphics[width=1in,height=1.25in,clip,keepaspectratio]{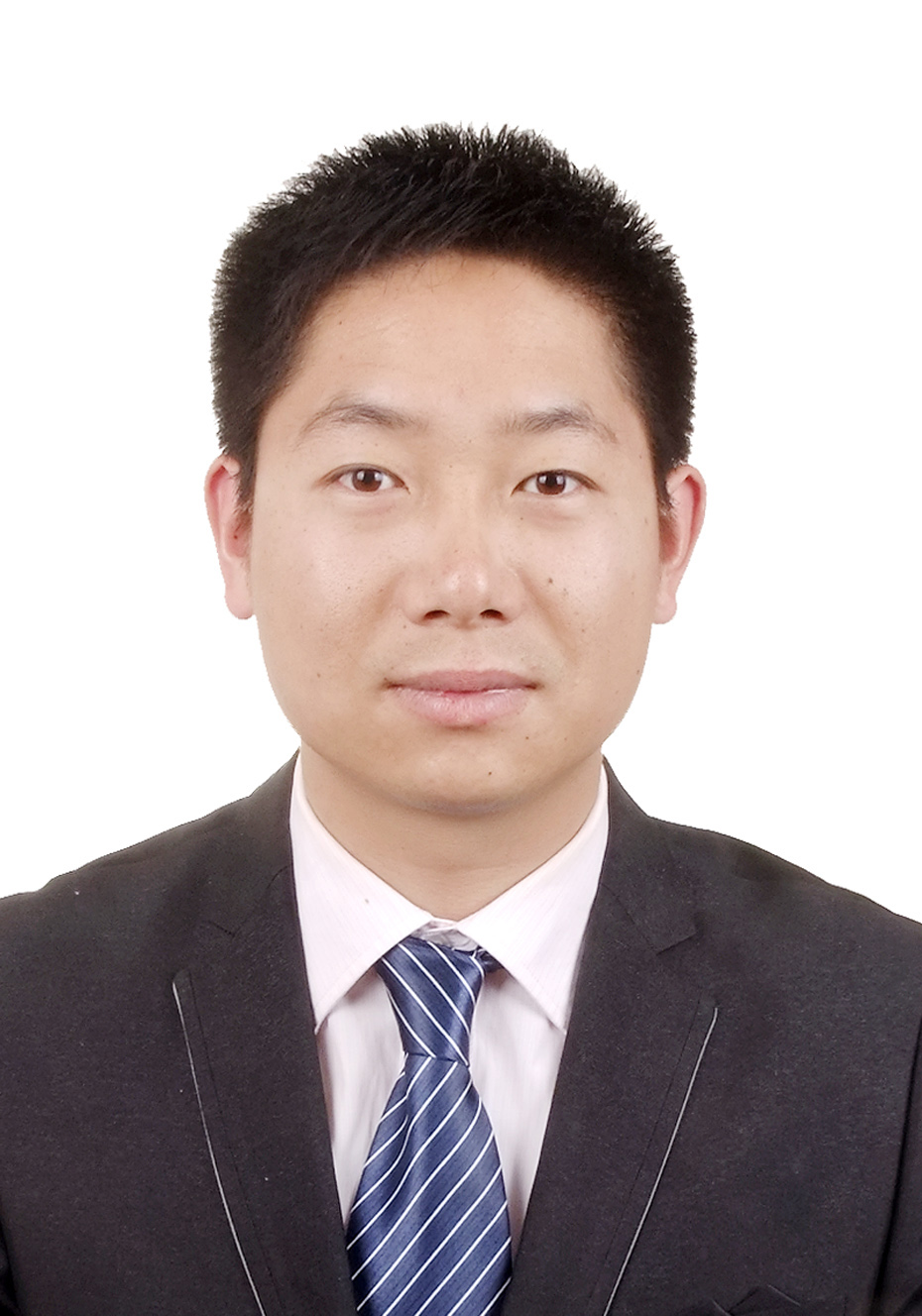}}]XUWEI WANG received the M.S. degree from the College of Civil Engineering and Architecture, Zhejiang University. He has 5 years working experience in computer vision field and obtains three authorized invention patents. He is currently a doctoral student majoring in Computer Science and Technology, Zhejiang University. His research interests include AI, Deep Learning and Computer Vision.
\end{IEEEbiography}

\begin{IEEEbiography}[{\includegraphics[width=1in,height=1.25in,clip,keepaspectratio]{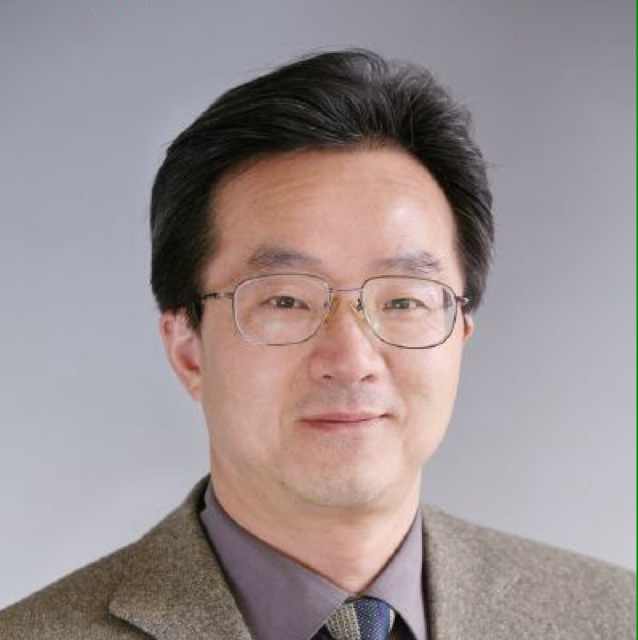}}]QING LI is currently a Professor with China Jiliang University. He was the dean of the School of Mechanical and Electrical Engineering, China Jiliang University and now he is an executive director of China's metrological testing association. He was selected as a famous teacher of the national "Ten-Thousand Talents Program" in 2017. His current research direction is dynamic measurement and control, sensing technology. Presided over the completion of more than 60 science and technology projects.
\end{IEEEbiography}

\begin{IEEEbiography}[{\includegraphics[width=1in,height=1.25in,clip,keepaspectratio]{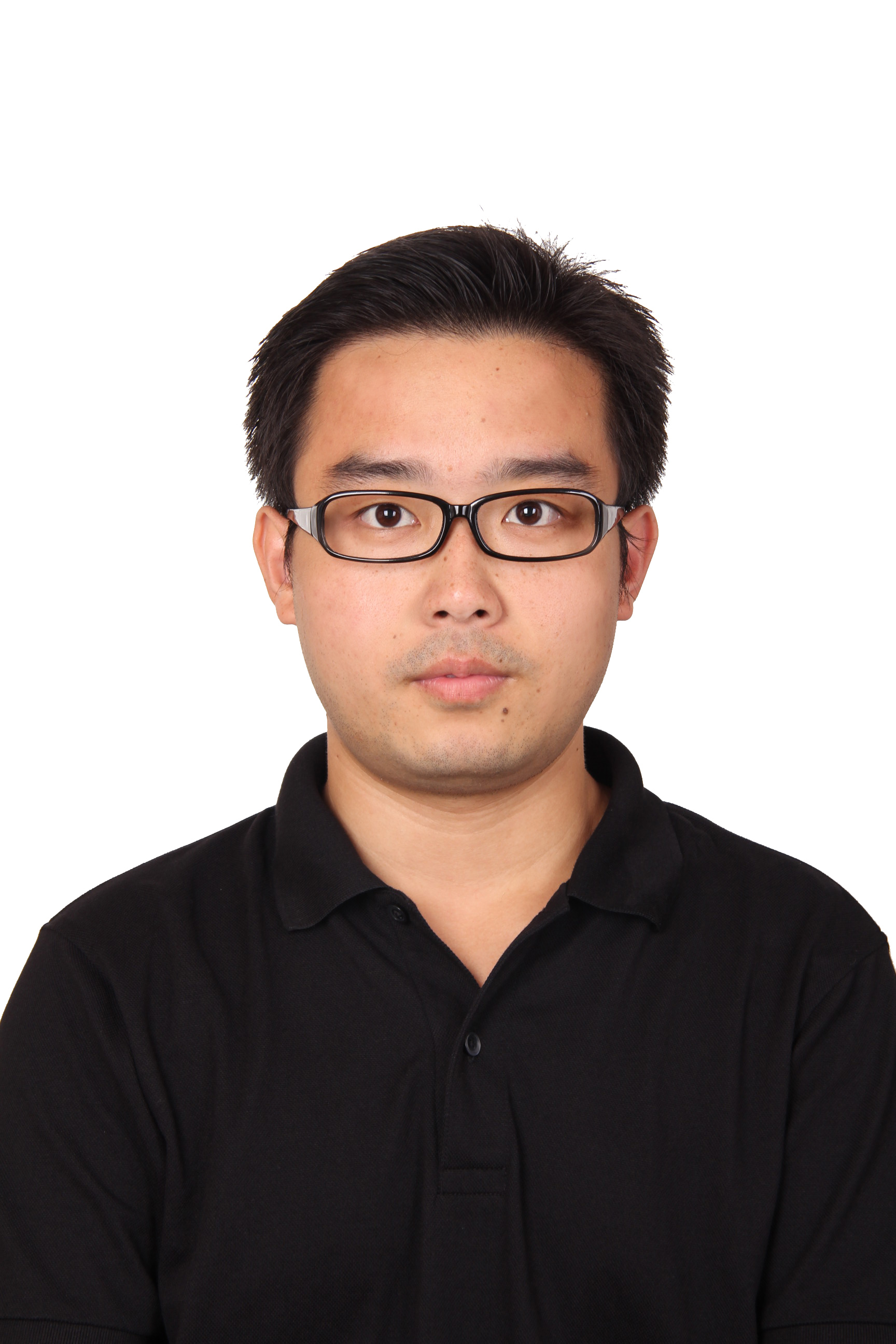}}]SHENGYAO JIA received the B.E. degree in measurement and control technology and instrumentation and the M.S. degree in detection technology and automation from China Jiliang University, Hangzhou, China, in 2008 and 2011, respectively, and the Ph.D. degree in agricultural electrification and automation from Zhejiang University, Hangzhou, China, in 2015. He is a lecturer with China Jiliang University. His current research interests include energy harvesting systems, sensors and measuring technology and embedded system.
\end{IEEEbiography}

\begin{IEEEbiography}[{\includegraphics[width=1in,height=1.25in,clip,keepaspectratio]{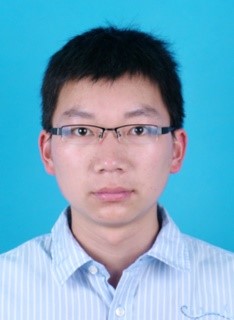}}]RENYUAN TONG was graduated from East China Normal University. He is now an associate professor in College of Mechanical and Electrical Engineering, China Jiliang University. His research interest is detection technology.
\end{IEEEbiography}

\EOD

\end{document}